\documentclass[letter, twocolumn]{jpsj3} 

\usepackage{epstopdf}
\title{Upper critical field and de Haas-van Alphen oscillations in KOs$_2$O$_6$ measured in a hybrid magnet}

\author{Taichi \textsc{Terashima}$^{1}$, Nobuyuki \textsc{Kurita}$^{1}$, Atsushi \textsc{Harada}$^{1}$, Kota \textsc{Kodama}$^{1}$, Jun-ichi \textsc{Yamaura}$^{2}$, Zenji \textsc{Hiroi}$^{2}$, Hisatomo \textsc{Harima}$^{3}$, Shinya \textsc{Uji}$^{1}$}

\inst{$^{1}$National Institute for Materials Science, Tsukuba, Ibaraki 305-0003, Japan\\
$^{2}$Institute for Solid State Physics, University of Tokyo, Kashiwa, Chiba 277-8581, Japan\\
$^{3}$Department of Physics, Graduate School of Science, Kobe University, Kobe, Hyogo 657-8501, Japan}

\abst{Magnetic torque measurements have been performed on a KOs$_2$O$_6$ single crystal in magnetic fields up to 35.3 T and at temperatures down to 0.6 K.  The upper critical field is determined to be $\sim$30 T.  De Haas-van Alphen oscillations are observed.  A large mass enhancement of (1+$\lambda$) = $m^* / m_{band}$ = 7.6 is found.  It is suggested that, for the large upper critical field to be reconciled with Pauli paramagnetic limiting, the observed mass enhancement must be of electron-phonon origin, including electron-rattling-mode interactions, for the most part.}

\kword{pyrochlore oxides, KOs$_2$O$_6$, upper critical field, de Haa-van Alphen effect}

\begin{document}
\maketitle

The $\beta$-pyrochlore osmium oxides AOs$_2$O$_6$ (A = K, Rb, and Cs) exhibit superconductivity with the transition temperatures $T_c$ of 9.6, 6.3, and 3.3 K, respectively.\cite{Yonezawa04JPCM, Yonezawa04JPSJ_Rb, Yonezawa04JPSJ_Cs, Hiroi05JPSJ_erratum}  In a recent paper,\cite{Nagao09JPSJ} Hiroi and coworkers have argued that pairing in these compounds is mediated by a low-energy rattling mode.  A$^+$ ions in AOs$_2$O$_6$ are enclosed in much bigger cages formed by OsO$_6$ octahedra and vibrate with large amplitudes in an anharmonic potential.\cite{Kunes04PRB}  Such vibrations are called rattling.  This unique variant of electron-phonon superconductivity deserves further detailed studies.

We here report magnetic torque measurements on KOs$_2$O$_6$.  We focus on two subjects.  The first is the upper critical field $B_{c2}$.  We confirm a large upper critical field at zero temperature $B_{c2}(0) \sim 30$ T as previously reported.\cite{Shibauchi06PRB, Ohmichi06JPSJ}  For comparison, the paramagnetic critical field $B_{po}$ is 18 T if evaluated as $B_{po} = 1.84 T_c$ ($B_{po}$ in Tesla and $T_c$ in Kelvin) as usual.  To resolve this apparent contradiction, a previous interpretation invoked noncentrosymmetry of the crystal structure.\cite{Shibauchi06PRB, Schuck06PRB}  However, it is now clear that the crystal structure remains centrosymmetric down to low temperatures below $T_c$.\cite{Yamaura09SSC, Hasegawa08PRB, Sasai10JPCM, Yamaura10JPSJ}  Accordingly, we reexamine if and how the large $B_{c2}$(0) can be reconciled with Pauli limiting.  The other is the de Haas-van Alphen (dHvA) effect.  We have observed dHvA oscillations in KOs$_2$O$_6$ for the first time.  From comparison with band structure calculations, we find an unusually large mass enhancement for the basically electron-phonon superconductor, confirming previous analyses of specific heat data.\cite{Hiroi07PRB, Nagao09JPSJ}  Since the specific heat was measured only up to 14 T, where $T_c$ is still 5.2 K, the Sommerfeld coefficient $\gamma$ at low temperatures could not be determined directly but was estimated on the basis of a few assumptions including an unconventional temperature dependence of the lattice contribution.  Therefore the present direct confirmation is invaluable.

\begin{figure}
\begin{center}
\includegraphics[width=8cm]{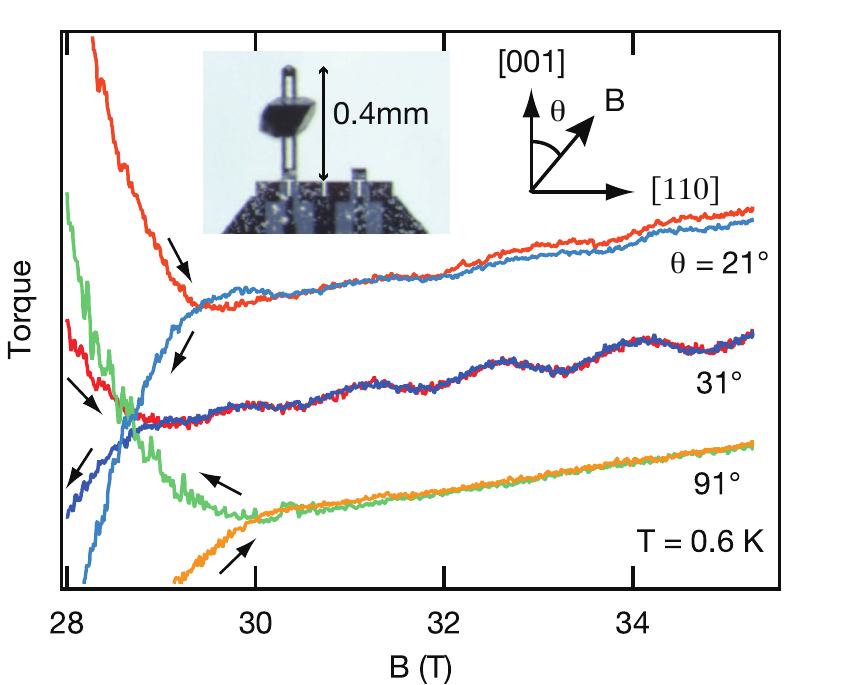}
\end{center}
\caption{(Color online) Magnetic torque of KOs$_2$O$_6$ at 0.6 K for three different field directions.  Both field up and field down sweeps are shown.  The left inset shows the single crystal sample mounted on a microcantilever, and the right the definition of the field angle $\theta$.}
\label{f1}
\end{figure}

The single crystal used in the present study was prepared as described in ref.~\citen{Hiroi07PRB}, where a residual resistivity ratio of about 300 was reported.  Magnetic fields up to 35.3 T were produced by the hybrid magnet installed at the Tsukuba Magnet Laboratory of the NIMS.\cite{TML}  Low temperatures down to 0.6 K were generated by a $^3$He refrigerator.  Magnetic torque was detected by using a piezoresistive microcantilever (Fig.~\ref{f1}).  The field direction $\theta$ is measured from [001] toward [110] (Fig.~\ref{f1}).  The band structure calculations were performed within the local density approximation using a full potential LAPW (FLAPW) method.\cite{Hiroi07PhysC}  We used the program codes TSPACE\cite{Yanase1995} and KANSAI-06.  The obtained electronic band structure is very similar to that previously obtained for CsOs$_2$O$_6$.\cite{Terashima08PRB}  

\begin{figure}
\begin{center}
\includegraphics[width=8cm]{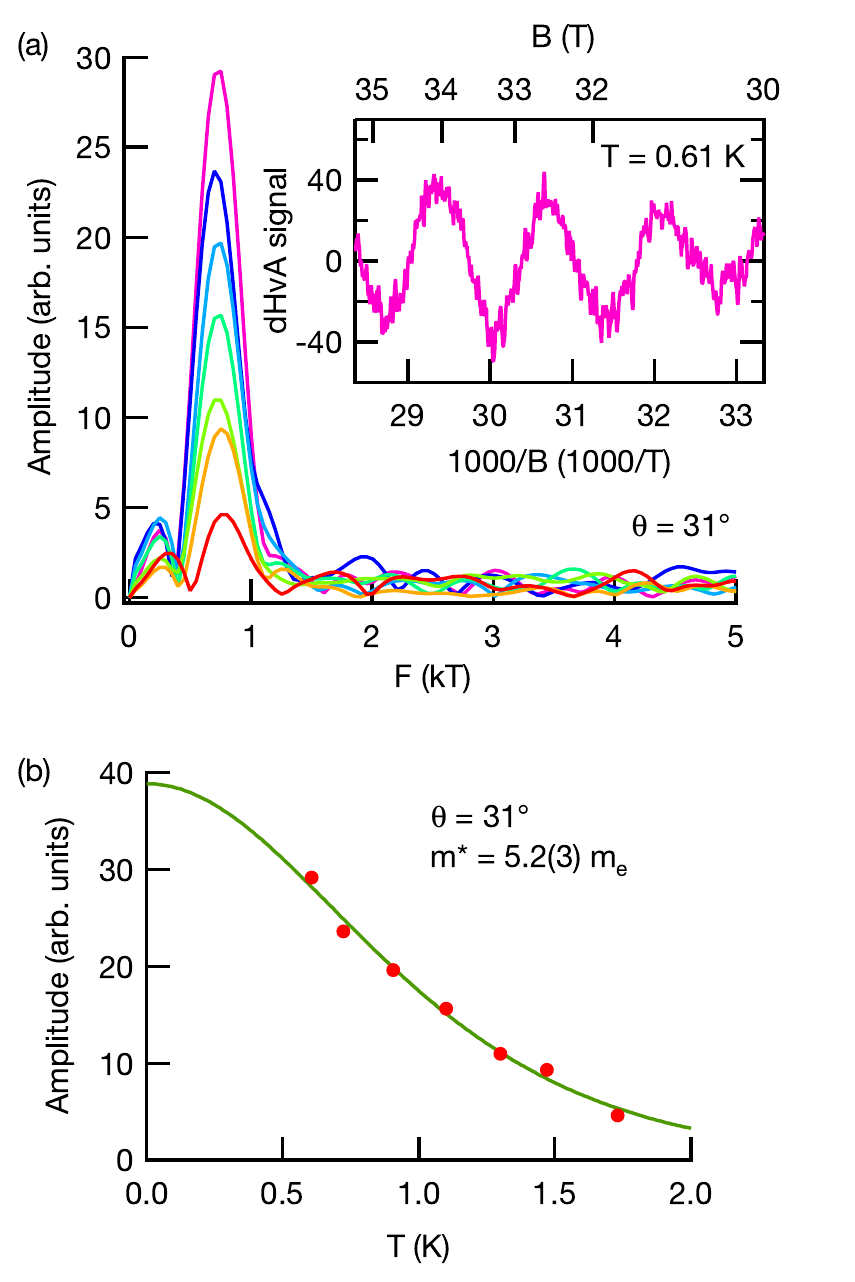}
\end{center}
\caption{(Color online) (a) The inset shows the dHvA oscillatory torque at $T$ = 0.61 K and $\theta$ = 31$^{\circ}$ obtained after smoothly-varying background was subtracted from the data in Fig.~\ref{f1}.  The up- and down-sweep data were averaged.  The main part shows Fourier transforms of oscillations (in 1/$B$) at $\theta$ = 31$^{\circ}$ for various temperatures.  $T$ = 0.61, 0.72, 0.91, 1.1, 1.3, 1.5, and 1.7 K with decreasing amplitude.  (b) Temperature dependence of the oscillation amplitude.  The solid curve is a fit to the Lifshitz-Kosevich formula, from which we deduce that $m^*$ = 5.2(3) $m_e$.}
\label{f2}
\end{figure}

\begin{figure}
\begin{center}
\includegraphics[width=8cm]{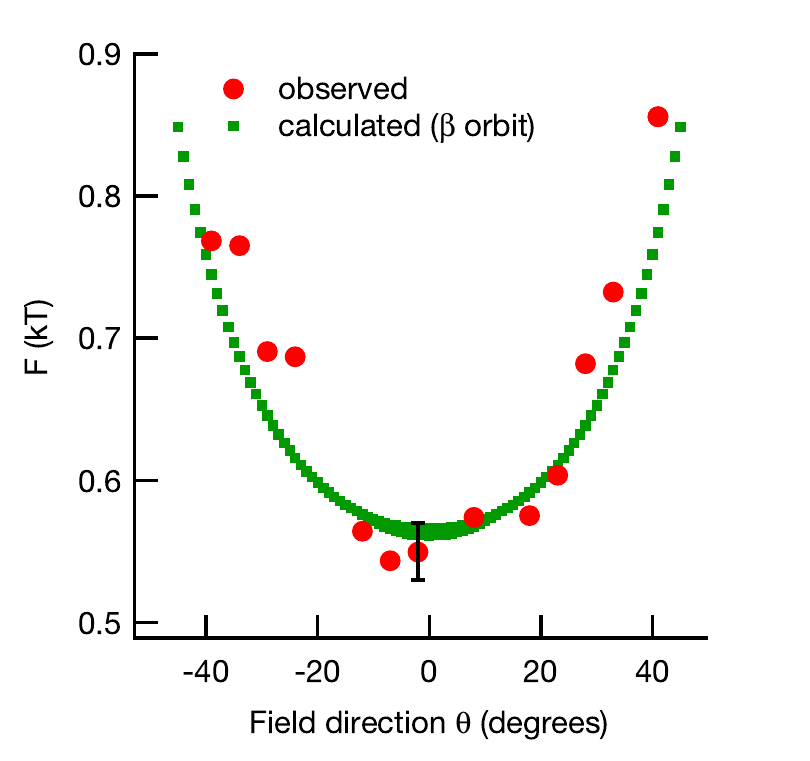}
\end{center}
\caption{(Color online) Angle dependence of the observed frequency compared to the calculated $\beta$ frequency.}
\label{f3}
\end{figure}

\begin{figure}
\begin{center}
\includegraphics[width=8cm]{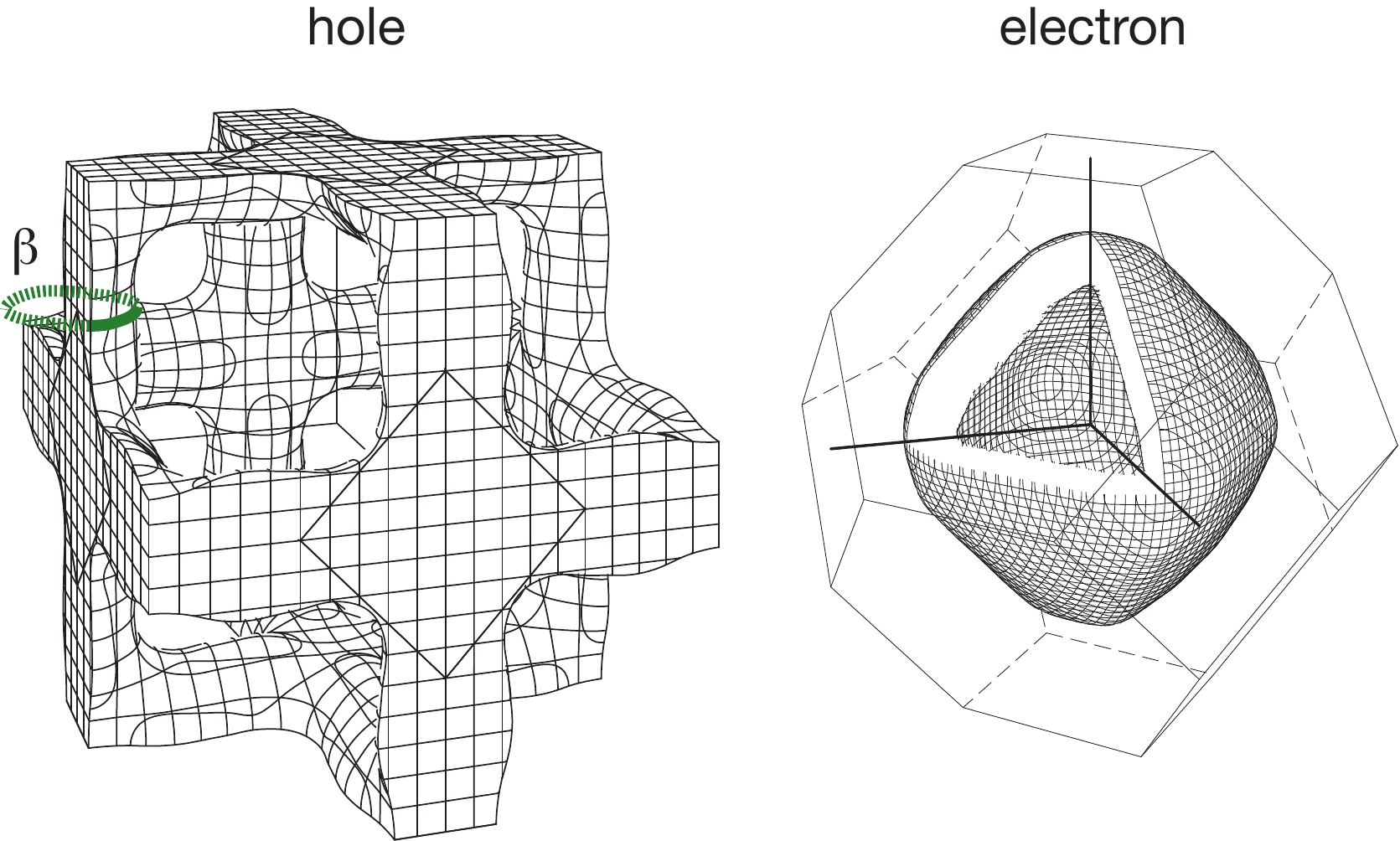}
\end{center}
\caption{(Color online) Calculated Fermi surface of KOs$_2$O$_6$.  The experimentally observed $\beta$ orbit is on the hole sheet.}
\label{f4}
\end{figure}

Figure~\ref{f1} shows magnetic torque measured at 0.6 K as a function of field for three field directions, where both field up and field down sweeps were made.  The up- and down-field traces diverge below 30 T for $\theta$ = 21 and 91$^{\circ}$.  For $\theta$ = 31$^{\circ}$ the divergence occurs at a slightly lower field, i.e., 29 T.  We identify these fields with $B_{c2}(0)$ and conclude that $B_{c2}(0) \sim 30$ T, assuming $B_{c2}(0) \approx B_{c2}(0.6 \mathrm{K})$.\cite{Birr}  This estimation is close to $\sim$33 and 30.6 T reported in refs.~\citen{Shibauchi06PRB} and \citen{Ohmichi06JPSJ}, respectively.  The smaller field value observed for $\theta$ = 31$^{\circ}$ may indicate the anisotropy of $B_{c2}$, which can exist even in a cubic material if the Fermi surface is nonspherical.\cite{Hohenberg67PR}

dHvA oscillations are already evident for $\theta$ = 21 and 31$^{\circ}$ above $B_{c2}$ even before smoothly-varying background is subtracted.  The dHvA oscillatory torque at $\theta$ = 31$^{\circ}$ obtained after the background subtraction and the corresponding Fourier transform are shown Fig.~\ref{f2}(a).  A single peak appears at a dHvA frequency $F$ = 0.73 kT in the transform.  The angle dependence of $F$ is shown in Fig.~\ref{f3}.  $F$ = 0.54(2) kT for $B \parallel $ [001].  The observed frequency branch is in an excellent agreement with the calculated $\beta$ branch, which was named after the corresponding branch in CsOs$_2$O$_6$.\cite{Terashima08PRB}  The $\beta$ orbit is on the hole surface as shown in Fig.~\ref{f4}.

The temperature dependence of the dHvA oscillation amplitude measured at $\theta$ = 31$^{\circ}$ is shown in Fig.~\ref{f2}.  The effective mass $m^*$ is estimated from a fit to the Lifshitz-Kosevich formula\cite{Shoenberg84} to be 5.2(3) $m_e$, $m_e$ being the free electron mass.  According to our band structure calculations, the band mass $m_{band}$ is 0.68$m_e$.  Thus the mass enhancement (1+$\lambda$) is estimated to be (1+$\lambda$) = $m^* / m_{band}$ = 7.6, which is very close to the value of 7.3 estimated from the specific heat data.\cite{Nagao09JPSJ}  In the case of CsOs$_2$O$_6$, the mass enhancement is 4.2 for the $\beta$ orbit.\cite{Terashima08PRB}.  Thus the mass enhancement for the $\beta$ orbit is increased by 81\% from CsOs$_2$O$_6$ to KOs$_2$O$_6$.  This is also consistent with a 94\% increase in the specific heat mass enhancement [(1+$\lambda$) = 3.76 from the specific heat in CsOs$_2$O$_6$].\cite{Nagao09JPSJ}

Three points should be noted here.  Firstly, the observed mass enhancement of 7.6 is unusually large when compared to values found in other superconductors with relatively high $T_c$'s.  dHvA measurements on LuNi$_2$B$_2$C ($T_c$ = 16 K) and MgB$_2$ ($T_c$ = 38 K) have observed, respectively, the mass enhancements (1+$\lambda$) of 1.3--3.7\cite{Bergk08PRL} (or 1.16--3.02\cite{Isshiki08PRB}) and those of 1.31--2.2\cite{Carrington03PRL}, depending on orbits.  The mass enhancements of 2--3 have been found for the A15 compounds Nb$_3$Sn ($T_c$ = 18 K) and V$_3$Si ($T_c$ = 17 K) and the Chevrel compounds SnMo$_6$S$_8$ and PbMo$_6$S$_8$ ($T_c$ = 12 K for both) from analyses of specific heat.\cite{Junod83PRB, Lachal84JLTP}

Secondly, the values of $m^*$ and $\gamma$ were estimated at very different magnetic fields: $m^*$'s were estimated at 9.3 and 32 T for CsOs$_2$O$_6$ and KOs$_2$O$_6$, respectively, while $\gamma$'s were estimated for zero field.  The observed consistency among these data indicates that the mass enhancement is insensitive to the magnetic field. 

Thirdly, the observed consistency also suggests that the mass enhancement is fairly homogeneous over the Fermi surface.  At least, the mass enhancement does not vary as it does in LuNi$_2$B$_2$C, where it varies from 1.3 to 3.7\cite{Bergk08PRL} (or 1.16 to 3.02\cite{Isshiki08PRB}) depending on orbits.

No other dHvA frequency than $\beta$ has been observed in the present study.  This can basically be explained by results of the band structure calculations: Namely, band masses associated with other possible dHvA frequencies are heavier than the mass of $\beta$.  Because of relatively high measurement temperatures ($T \geqslant 0.6$ K), dHvA oscillations are rapidly suppressed as effective masses increase.  In addition, since amplitude of dHvA torque oscillation is proportional to d$\ln F$/d$\theta$, it is difficult to detect dHvA frequency branches with little angular dependence, like the $\gamma$ and $\delta$ branches in CsOs$_2$O$_6$,\cite{Terashima08PRB} in torque measurements.  

Having determined the mass enhancement in KOs$_2$O$_6$, we now return to the large $B_{c2}$.  As one goes from CsOs$_2$O$_6$ to KOs$_2$O$_6$, $B_{c2}$(0) increases from 1.4\cite{Terashima08PRB} to $\sim$30 T.  Since $T_c$ increases only by a factor of three, this factor-of-21 increase may appear surprising at first glance.  However, this increase is consistent with an expected increase in the orbital critical field $B_{c2}^*(0)$.  In the clean limit of an isotropic single-band model, $B_{c2}(0)^* \propto \gamma^2 T_c^2 = \gamma_{band}^2 (1+\lambda)^2 T_c^2$, where $\gamma_{band}$ is an unrenormalized band value.\cite{Werthamer66PRB, Orlando79PRB, Decroux82Book}  Using $(1+\lambda)$ = 4.2 and 7.6 determined from the dHvA measurements and $\gamma_{band}$ = 11.1 and 9.6 mJK$^{-2}$mol$^{-1}$ for CsOs$_2$O$_6$ and KOs$_2$O$_6$,\cite{Nagao09JPSJ, Terashima08PRB} respectively, we can estimate the ratio of $B_{c2}^*(0)$ between the two compounds to be 21.  The agreement between the observed and estimated factors indicates that $B_{c2}(0)$ is basically determined by orbital effects in these compounds with only minor influence of spin effects, as previously proposed in refs.~\citen{Shibauchi06PRB} and \citen{Bruhwiler06PRB}.  Numerical estimations can be made from the formula $B_{c2}^*(0) = -0.73 \mathrm{d}B_{c2} / \mathrm{d}T |_{T=T_c} T_c$.\cite{Werthamer66PRB}  Using experimental values of $-\mathrm{d}B_{c2} / \mathrm{d}T |_{T=T_c}$ = 0.44 and 3.61 T/K,\cite{Nagao09JPSJ, Hiroi07PRB, Bruhwiler06PRB} we have $B_{c2}^*(0)$ = 1.0 and 25 T for CsOs$_2$O$_6$ and KOs$_2$O$_6$, respectively.  The fact that $B_{c2}(0) > B_{c2}^*(0)$ may be accounted for by multi band (KOs$_2$O$_6$ is a two-band system) and Fermi surface effects as proposed in ref.~\citen{Shibauchi06PRB}.

Next we consider how the large $B_{c2}(0)$ can be reconciled with the small $B_{po}$.  When electron-phonon and electron-electron interactions exist, the paramagnetic critical field $B_p$ is modified from the BCS one $B_{po}$: $B_p$ = $B_{po} (1+\lambda) / S$, where $S$ is the Stoner enhancement factor of the spin susceptibility.\cite{Orlando81PRL, Decroux82Book}

In a previous work,\cite{Terashima08PRB} we have determined the product $Sg$, where $g$ is the $g$ factor, to be 9.4 for the $\beta$ orbit in CsOs$_2$O$_6$.  If we assume $g$ = 2, we have $S$ = 4.7.  Using this value and $(1+\lambda)$ = 7.6, we obtain $B_p$ = 29 T for KOs$_2$O$_6$, which is still insufficient to explain the observed $B_{c2}(0)$.  In this argument, $S$ may have been overestimated, because, in general, electrons circuiting a small orbit that crosses the Brillouin zone boundary several times as the present $\beta$ orbit may have a $g$ value much larger than 2.\cite{Shoenberg84}

We therefore switch to values of $S$ determined from bulk magnetic susceptibility measurements.\cite{Hiroi07PRB, Nagao09JPSJ}  $S$ = 3.1 for CsOs$_2$O$_6$ in ref.~\citen{Terashima08PRB} whereas $S$ = 1.3 and 1.2 for CsOs$_2$O$_6$ and KOs$_2$O$_6$, respectively, in ref.~\citen{Nagao09JPSJ}.  In the former, a possible contribution of the Van Vleck term was neglected, while in the latter the Van Vleck term was assumed to be the same as that in Cd$_2$Re$_2$O$_7$, which was determined from the $K$ (Knight shift)-$\chi$ plot in Cd NMR measurements,\cite{Vyaselev02PRL} and was subtracted from the measured total susceptibilities.  With $S$ = 3.1 or 1.2, we have $B_p$ = 43 or 112 T, respectively.

In order to see a combined effect of $B_{c2}^*(0)$ and $B_p$, we may resort to a dirty-limit formula, $B_{c2}(0)^{-2} = B_{c2}^*(0)^{-2} + 2 B_p^{-2}$.\cite{Orlando81PRL, Decroux82Book}  If we assume $B_{c2}^*(0)$ = 33 T (rather arbitrarily), we have $B_{c2}(0)$ = 22 or 30 T for $B_p$ = 43 or 112 T, respectively.  Although impurity spin-orbit scattering may increase thus calculated $B_{c2}(0)$ values, such an effect should be small for the present high-quality crystal.  It is clear that $B_p$ = 112 T that was derived with $S$ = 1.2 is more appropriate.  We therefore conclude that, to be consistent with the observed large $B_{c2}(0)$, the Stoner enhancement factor must be close to 1 in KOs$_2$O$_6$ as estimated in ref.~\citen{Nagao09JPSJ}.  This immediately implies that the observed mass enhancement $(1+\lambda)$ = 7.6 is due to electron-phonon interactions, including electron-rattling-mode ones, for the most part: Let us assume $(1+\lambda) = (1+\lambda_{ep})(1+\lambda_{ee})$, where $\lambda_{ep}$ and $\lambda_{ee}$ are an electron-phonon and an electron-electron coupling constant, respectively.  Effects of electron-rattling-mode interactions are included in the former.  In general, $S/(1+\lambda_{ee}) > 1$.\cite{Wilkins80inbook}  Therefore, if $S$ = 1.2, $(1+\lambda_{ee})  < 1.2$, and hence $(1+\lambda_{ep})  > 6.3$.

In conclusion, we have performed magnetic torque measurements on KOs$_2$O$_6$.  We have confirmed that the upper critical field amounts to $\sim$30 T.  We have observed de Haas-van Alphen oscillations arising from the $\beta$ orbit on the hole Fermi surface.  The size and angular dependence of the $\beta$ orbit are in excellent agreement with the band structure calculation.  A large mass enhancement of (1+$\lambda$) = $m^* / m_{band}$ = 7.6 has been found.  Considering how the large $B_{c2}$ can be reconciled with Pauli limiting, we have concluded that the Stoner enhancement factor must be close to one and hence that the observed mass enhancement must be of electron-phonon origin, including electron-rattling-mode interactions, for the most part.


\begin{thebibliography}{10}
\bibitem{Birr}
Exactly, the fields thus determined are the irreversibility fields $B_{irr}$.  The difference between $B_{irr}$ and $B_{c2}$ may be appreciable in highly two-dimensional superconductors.  However, KOs$_2$O$_6$ is a three dimensional material, and hence the difference between $B_{irr}$ and $B_{c2}$ in the temperature range of the present study can safely be assumed to be vanishingly small.  Indeed, none of the measured torque vs. magnetic field curves shows a kink or bend above $B_{irr}$ that may indicate a separate $B_{c2}$.
\end{thebibliography}

\end{document}